\documentstyle[12pt]{article}
\begin{document}
\def\thefootnote{\fnsymbol{footnote}}
\begin{flushright}
KANAZAWA-02-02  \\ 
February, 2002
\end{flushright}
\vspace*{2cm}
\begin{center}
{\LARGE\bf Neutrino mass due to the neutrino-gaugino
 mixing}\footnote[1]{
Talk given at Workshop in the 
Corfu Summer Institute CORFU2001.}\\
\vspace{1 cm}
{\Large  Daijiro Suematsu}
\footnote[2]{e-mail:suematsu@hep.s.kanazawa-u.ac.jp}
\vspace {0.7cm}\\
$^\ast${\it Institute for Theoretical Physics, Kanazawa University,\\
        Kanazawa 920-1192, JAPAN}
\end{center}
\vspace{2cm}
{\Large\bf Abstract}\\  
We study a possibility to explain neutrino masses and mixings 
based on supersymmetry. If we introduce a flavor diagonal but 
generation dependent extra U(1) gauge interaction at a TeV region, 
we can obtain masses and mixings of neutrinos required for the 
explanation of both solar and atmospheric neutrinos. In this model, 
differently from the usual bilinear $R$-parity violating scenario, 
the neutrino mass degeneracy can be resolved at a tree level by the
neutrino-gaugino mixing caused by an $R$-parity violation. The model 
is straightforwardly extended to include the quark sector by 
introducing an anomalous U(1) which can be used for the 
Froggatt-Nielsen mechanism.
\newpage
\section{Introduction}
Recent observations of the solar neutrinos \cite{solar} and the 
atmospheric neutrinos \cite{atmos} 
at Super Kamiokande suggest that neutrinos have small masses and
there are large flavor mixings in the lepton sector. 
These features are quite different from the ones of the quark sector and
it may be a clue to
the developement of a unification picture of quarks and leptons
for us to ask "what is the origin of these differences?".

The mixing in the lepton sector is represented by the so-called 
MNS matrix which is defined by $V^{\rm MNS}=U_\ell^\dagger U_\nu$,
where $U_\ell$ and $U_\nu$ are the mixing matrices for charged leptons
and neutrinos, respectively.
Thus the origin of large mixings in the lepton sector exists 
in either $U_\ell$ or $U_\nu$.
A famous example of possible mass matrices to realize large mixings
is a democratic form, which takes the form such as 
\begin{equation}
m\left(\begin{array}{ccc}
1 & 1 & 1 \\ 1 & 1 & 1 \\ 1 & 1 & 1 \\
\end{array}\right).
\end{equation}
However, this matrix has a serious problem to apply it to the neutrino sector. 
It has only a nonzero eigenvalue and then both the solar and atmospheric 
neutrino problems cannot be explained. It needs a suitable deviation from the
exact democratic form to overcome this fault. 
It may be worthy to study how to generate an appropriately deviated
form from it based on a certain physical principle.
We study this problem on the basis of {\it supersymmetry} and an {\it
extra U(1) symmetry} in a TeV region. The former one is related to the
gauge hierarchy problem and the latter one often appears in the effective
theory of superstring. So this kind of study is considered to be sufficiently 
motivated.

In the supersymmetric theory there generally exists a discrete symmetry
called $R$-parity, which is defined by $R_p=(-1)^{3B+L+2S}$ where $B$ and $L$
are a baryon number and a lepton number, respectively and $S$ stands for 
a spin.
Under this symmetry particles in the standard model (SM) have $R_p=+1$ and
their superpartners have $R_p=-1$. Thus the SM particles cannot mix with their
superpartners without an $R$-parity violation. 
In the minimal supersymmetric SM (MSSM) 
there are neutral fermions with $R_p=-1$ in addition to the neutrinos.
They are called neutralinos and contain
$(\lambda_{W_3}, \lambda_Y, \tilde H_1^0, \tilde H_2^0)$.
$\lambda_{W_3}$ and $\lambda_Y$ are gauginos for SU(2)$_L\times$
U(1)$_Y$ gauge symmetry and 
$\tilde H_1^0$ and $\tilde H_2^0$ stand for Higgsinos. 
If $R_p$ is not broken, neutrinos and neutralinos cannot mix.
However, if there are $R_p$ violations such as an explicit breaking due to 
the bilinear $R$-parity violating terms $\epsilon_\alpha L_\alpha H_2$ 
in superpotential or a spontaneous breaking due to the 
nonzero vacuum expectation values (VEVs) of sneutrinos, 
mass mixings ${\cal M}$ among 
neutrinos and neutralinos appear in the form such as \cite{rparity1,rparity2}
\begin{equation}
(\nu_e,\nu_\mu, \nu_\tau)\left(
\begin{array}{ccc}
\sqrt 2g_2\langle\tilde\nu_e\rangle &\sqrt 2g_1\langle\tilde\nu_e\rangle 
& \epsilon_e \\
\sqrt 2g_2\langle\tilde\nu_\mu\rangle 
&\sqrt 2g_1\langle\tilde\nu_\mu\rangle & \epsilon_\mu \\
\sqrt 2g_2\langle\tilde\nu_\tau\rangle 
&\sqrt 2g_1\langle\tilde\nu_\tau\rangle & \epsilon_\tau \\
\end{array}\right)
\left( \begin{array} {c} -i\lambda_{W_3} \\ -i\lambda_Y \\ \tilde H_2^0 \\
\end{array} \right).
\label{eqa}
\end{equation}
A study of the scalar potential indicates that $\langle\tilde\nu_\alpha\rangle$
is proportional to $\epsilon_\alpha$. As a result of this feature, 
all column vectors 
of ${\cal M}$ are proportinal to each other.
Under the assumption that gaugino masses $M_1, M_2$ and a supersymmetric 
Higgs mass $\mu$ are 
much larger than 
$\langle\tilde\nu_\alpha\rangle$ and $\epsilon_\alpha$, we find that 
the neutrino mass matrix satisfies
$M_{ij}^\nu\propto \langle\tilde\nu_i\rangle\langle\tilde\nu_j\rangle$ as 
a result of the seesaw mechanism \cite{seesaw}. Unfortunately, the light 
neutrino mass matrix 
$M_{ij}^\nu$ obtained in this way has only one nonzero 
mass eigenvalue.
For the explanation of the atmospheric and solar neutrinos, we need 
a mass perturbation to resolve this mass degeneracy.  
A well-known example of such possibilities is an inclusion of one-loop effects 
and several works in this direction have been done by now \cite{loop1,loop2}.

In this talk we would like to propose an another possibility.
We consider a tree level solution to this problem
and study whether we can obtain large flavor mixings and 
also appropriate mass eigenvalues in the lepton 
sector in such a scenario.

\section{A simple Model}
We introduce an extra U(1) gauge symmetry to the MSSM, which 
is assumed to remain unbroken 
at a TeV region. This kind of extra U(1) has several interesting aspects
to take it seriously.
In fact, such a kind of symmetry often appears in the effective theory 
of superstring
and it can also present a natural solution to the $\mu$-problem \cite{mu}.
So there is a certain physical motivation to consider such a symmetry
in the model building.
It should be noted that there are experimental constraints on the 
extra U(1) from the precision measurements at LEP.
Its mass should be generally larger than 600GeV and its 
mixing angle $\xi$ with the ordinary 
$Z^0$ gauge boson has to be less than $10^{-3}$.

In the present consideration
we assume that the extra U(1) has flavor diagonal but generation dependent 
interactions such as
\begin{equation}
{\cal L}=\bar\nu_{\alpha}i\gamma^\mu(\partial_\mu-ig_Xq_\alpha A_\mu)
\nu_\alpha
+i\sqrt 2 g_X(\tilde\nu_\alpha^\ast\lambda_X q_\alpha\nu_\alpha
-\tilde\nu_\alpha\bar\lambda_Xq_\alpha\bar\nu_\alpha)+\cdots.
\label{eqb}
\end{equation} 
If sneutrinos get nonzero vacuum expectation values 
$\langle\tilde\nu_\alpha\rangle\not=0$ in eq.~(\ref{eqb}), 
neutrino-gaugino mixing appears through
$\sqrt 2g_Xq_\alpha\langle\tilde\nu_\alpha\rangle$, which can break the 
previously mentioned proportional relation among column vectors 
of the extended neutrino-neutralino mixing matrix corresponding to
(\ref{eqa}).\footnote{In this discussion we can safely ignore the
contributions from the mixings with Higgsinos. }
As a result, we can have two nonzero mass eigenvalues of neutrinos at 
the tree level \cite{gaugino}.
In the following discussion we study neutrino oscillation phenomena and other 
phenomenological features of this model.

We take charge assingments of this extra U(1) gauge symmetry 
for each generation of the leptons as follows,
\def\romth{I\hspace*{-0.8mm}I\hspace*{-0.8mm}I}
\def\romtw{I\hspace*{-0.7mm}I}
\begin{equation}
\ell_{L\alpha}~:~(q_I, q_I, q_{\romth}), \qquad 
\bar\ell_{R\alpha}~:~(-q_I, -q_I, -q_{\romth}),
\end{equation} 
where $\ell_{L\alpha}$ and $\bar\ell_{R\alpha}$ stand for 
SU(2)$_L$ doublet fields and its singlet charged ones, respectively.
We assume Higgs chiral superfields have no charge of the extra U(1). 
At this stage we do not 
identify the lepton flavor. If we assume that sneutrinos get VEVs such as
$\langle\tilde\nu_e\rangle=\langle\tilde\nu_\mu\rangle=
\langle\tilde\nu_\tau\rangle\equiv u\not=0$,
we have a neutrino-gaugino mass matrix
\begin{eqnarray}
&&{\cal M}=\left(\begin{array}{cc} 0 & m^T \\ m & M \\ \end{array}\right), \\
&&m=\left(\begin{array}{ccc} a_2 & a_1 & b \\ a_2 & a_1 & b \\
a_2 & a_1 & c \end{array}\right), \qquad
M=\left(\begin{array}{ccc} M_2 & 0 & 0 \\ 0 & M_1 & 0 \\
0 & 0 & M_X \end{array}\right), 
\end{eqnarray}
where $a_\ell={1\over \sqrt 2}g_\ell u$, $b=\sqrt 2g_Xq_Iu$ and 
$c=\sqrt 2g_Xq_{\romth}u$.
If elements of $m$ are much smaller than the ones of $M$, we can use 
the seesaw formula and then the neutrino mass matrix is found to be written 
as\footnote{
We can consider the same type texture of neutrino mass matrix 
in the ordinary seesaw framework with right-handed neutrinos 
by introducing suitable 
symmetries \cite{sterile}.} 
\begin{equation}
M^\nu=m^TM^{-1}m=\left(\begin{array}{ccc} 
m_0+\epsilon^2 & m_0+\epsilon^2 & m_0+\epsilon\delta \\ 
m_0+\epsilon^2 & m_0+\epsilon^2 & m_0+\epsilon\delta \\
m_0+\epsilon\delta & m_0+\epsilon\delta & m_0+\delta^2 \\ \end{array}\right),
\end{equation}
where $m_0={g_2^2u^2\over 2M_2}$, $\epsilon={g_Xq_Iu\over \sqrt{M_X}}$ and
$\delta={g_Xq_{\romth}u\over \sqrt{M_X}}$.
A diagonalization matrix $U$ of $M^\nu$ which is defined by
$U^\dagger M^\nu U= M^\nu_{\rm diag}$ is easily found to be
\begin{equation}
U=\left(\begin{array}{ccc}
{1\over\sqrt 2} & {1\over\sqrt 2}\cos\theta & {1\over\sqrt 2}\sin\theta \\
-{1\over\sqrt 2} & {1\over\sqrt 2}\cos\theta & {1\over\sqrt 2}\sin\theta \\
0 & -\sin\theta & \cos\theta \\
\end{array}\right).
\label{eqc}
\end{equation}
The mass eigenvalues and the mixing angle $\theta$ are written as
\begin{eqnarray}
&&m_1=0, \qquad 
m_{2,3}={1\over 2}\left[3m_0+2\epsilon^2
+\delta^2\mp\sqrt{(m_0+2\epsilon^2-\delta^2)^2
+8(m_0+\epsilon\delta)^2}\right], \\
&&\sin^22\theta={8(m_0+\epsilon\delta)^2\over
(m_0+2\epsilon^2-\delta^2)^2+8(m_0+\epsilon\delta)^2}.
\end{eqnarray}

If we assume that the charged lepton mass matrix is diagonal, 
a transition probability
for the neutrino oscillation $\nu_\alpha\rightarrow\nu_\beta$ can 
be written by using the elements of the mixing matrix $U$ as
\begin{equation}
{\cal P}_{\nu_\alpha\rightarrow\nu_\beta}(L)=\delta_{\alpha\beta}-4\sum_{i>j}
U_{\alpha i}U_{\beta i}U_{\alpha j}U_{\beta j}\sin^2
\left({\Delta m_{ij}^2\over 4E}L\right),
\end{equation}
where $\Delta m_{ij}^2=\vert m_i^2-m_j^2\vert$.
We summarize the contributions of the possible modes to the neutrino
oscillation  
$\nu_\alpha\rightarrow\nu_\beta$ in Table 1.
\begin{figure}[tb]
\begin{center}
\begin{tabular}{l|l|c|l}\hline
$(\alpha,\beta)$ & (i,~j) & 
$-4U_{\alpha i}U_{\beta i}U_{\alpha j}U_{\beta j}(\equiv{\cal A})$ 
& \\ \hline\hline
$(I, \romtw)$ & (1,~2) & $\cos^2\theta$  & (A) \\
        & (1,~3) & $\sin^2\theta$  & (B) \\
        & (2,~3) & $-\sin^2\theta\cos^2\theta$ & (C) \\ \hline
$(I, \romth)$& (2,~3) & $2\sin^2\theta\cos^2\theta$ & (D) \\ \hline
$(\romtw, \romth)$ & (2,~3) & $2\sin^2\theta\cos^2\theta$ & (E) \\ \hline
\end{tabular}
\vspace*{3mm}\\
Table 1~~Contributions to $\nu_\alpha\rightarrow\nu_\beta$.
\end{center}
\end{figure}
Here we assume that an inverse hierarchy and also an approximate degeneracy
between $\nu_2$ and $\nu_3$: $m_1 \ll m_2\sim m_3$.
In that case we find that the atmospheric neutrino requires 
$$ 2\times 10^{-3}{\rm eV}^2~{^<_\sim}~\Delta m_{12}^2\simeq\Delta m_{13}^2
~{^<_\sim}~6\times 10^{-3}{\rm eV}^2. $$
The solar neutrino is also found to require 
that $\Delta m_{23}^2$ should take a
value in a solution dependent way in the region such as
$$ 10^{-10}{\rm eV}^2~{^<_\sim}~\Delta m_{23}^2~{^<_\sim}~1.5\times
10^{-4}{\rm eV}^2. $$
Taking account of this, we find that we should identify the flavor 
$(I, \romtw, \romth)$ with $(\tau, \mu, e)$.
Then we can rewrite eq.~(\ref{eqc}) into the usual MNS mixing matrix as
\begin{equation}
U^{\rm MNS}=\left(\begin{array}{ccc}
0 & -\sin\theta & \cos\theta \\
-{1\over\sqrt 2} & {1\over\sqrt 2}\cos\theta & {1\over\sqrt 2}\sin\theta \\
{1\over\sqrt 2} & {1\over\sqrt 2}\cos\theta & {1\over\sqrt 2}\sin\theta \\
\end{array}\right).
\end{equation}
The atmospheric neutrino is explained by $\nu_\mu\rightarrow\nu_\tau$ 
which comes from the modes (A) and (B) in Table 1. 
On the other hand, the solar neutrino
is explained by
$\nu_e\rightarrow\nu_\mu$ (E) and $\nu_e\rightarrow\nu_\tau$ (D) whose 
total amplitude is ${\cal A}=\sin^22\theta$. 
Thus if we assume $\sin^22\theta\sim 10^{-2}$, the small mixing 
MSW solution (SMA) is realized. The large mixing MSW solution (LMA) and others 
(LOW and VO) are possible for the case of $\sin^22\theta\sim 1$.  
\input epsf 
\begin{figure}[tb]
\begin{center}
\epsfxsize=7.6cm
\leavevmode
\epsfbox{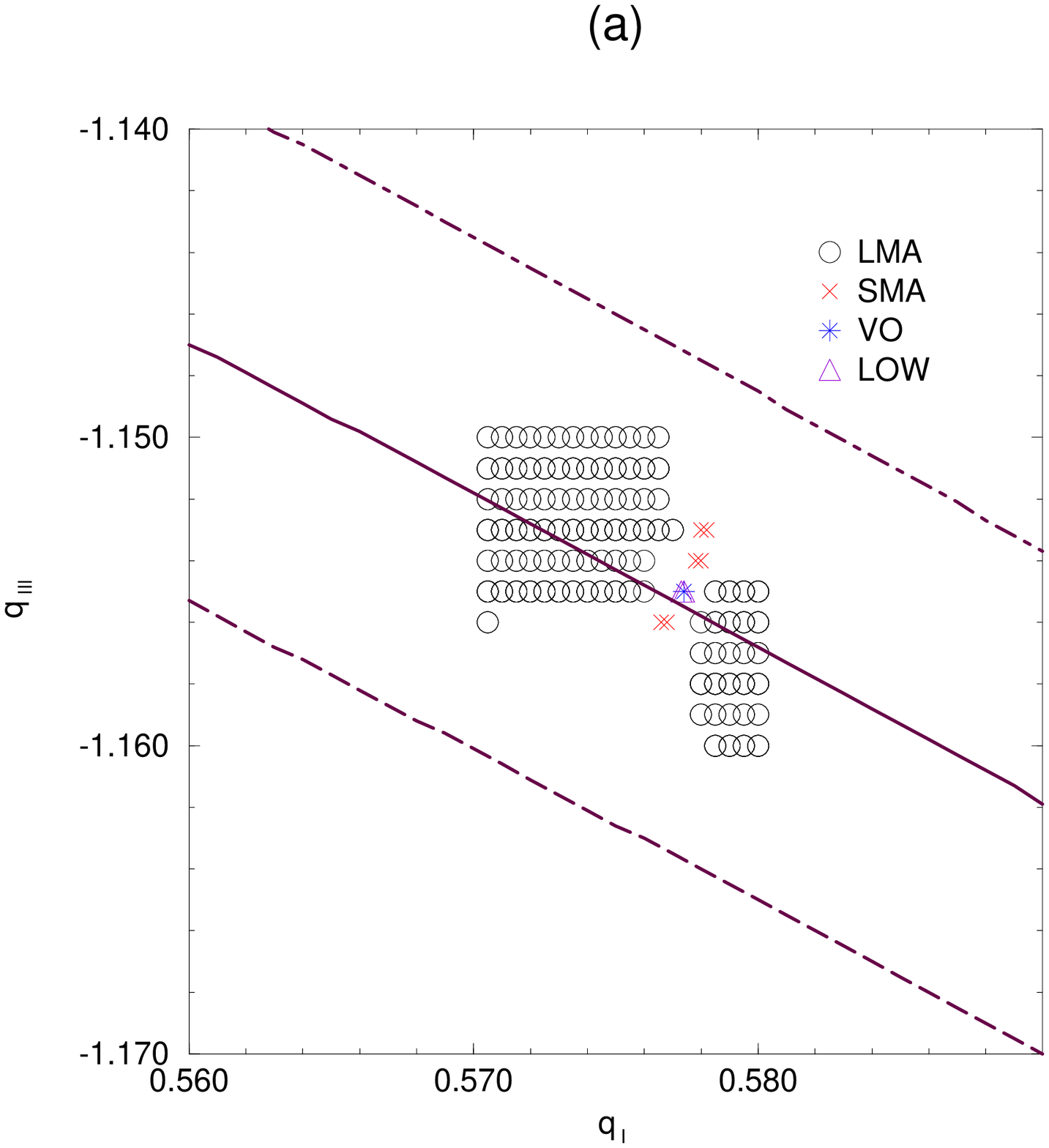}
\end{center}
\vspace*{-1cm}
{\footnotesize Fig. 1~~\  Scatter plots of the U(1)$_X$-charge
of neutrinos which can give a solution to the atmospheric and solar
neutrino problems. The dot-dashed, solid and dashed lines represent the
relation between $q_I$ and $q_{\romth}$ in the case that an addtional
singlet field takes the extra U(1) charges $Q=1.04,~1.05$ and 1.06, 
respectively.}
\end{figure}

We order several aspects of other neutrino phenomenology in this model here.\\
(1)~CHOOZ constraint \cite{chooz}~:~It is related to the neutrino oscillations 
$\nu_e\rightarrow\nu_{\mu,\tau}$ with 
$\Delta m_{12}^2$ and $\Delta m_{13}^2$ which are relevant to 
the atmospheric neutrinos.
Their ampitudes are propotinal to $U_{e1}^{\rm MNS}$ in this model. 
However, it is zero
and then the CHOOZ constraint is trivially satisfied.\\
(2)~Neutrinoless double $\beta$-decay \cite{beta}~:~ 
The effective mass parameter relevant to the neutrinoless double $\beta$
decay can be estimated as 
\begin{equation}
\vert m_{ee}\vert=\vert\sum_j\vert U_{ej}\vert^2 e^{i\phi_j}m_j\vert
=m_2\sin^2\theta+m_3\cos^2\theta\sim m_3,
\end{equation} 
where we assume $\phi_j=0$ and use the fact $m_2\sim m_3$. 
Then we have $\vert m_{ee}\vert\sim 0.04-0.08$~eV independently of 
the value of $\sin\theta$. This value seems to be a promising one for 
near future experiments.\\
(3)~Effects of the mode (C)~:~This corresponds to 
the oscillation $\nu_\mu\rightarrow\nu_\tau$, 
which is irrelevant to the atmospheric neutrino and also the 
short baseline experiments 
because of the smallness of $\Delta m_{23}^2$. However, if 
$\sin^22\theta\simeq 1$
and $\Delta m_{23}^2\sim 10^{-4}{\rm eV}^2$ which correspond to the LMA
solution are satisfied,
it can effectively contribute to the deviation from the simple two flavor 
oscillation scenario at $L~{^>_\sim}~2000$~km.

\section{Various issues of the model}
In this section we will briefly comment on the issues 
which are crucial for the model to be a phenomenologically viable model.

(1)~Realization of oscillation parameters\\
We have $M_A, g_A~(A=2,1,X)$, $q_\alpha$ and $u$ as the parameters 
in our model.
If we assume a gauge coupling unification and a universal gaugino 
mass $M_0$ at $M_{\rm GUT}$, gaugino masses at a low energy 
region are determined by renormalization group equations as
\begin{equation}
M_2(\mu)={M_0\over g_U^2}g_2^2(\mu), \qquad M_{1,X}(\mu)={M_0\over 
g_U^2}g_{1,X}^2(\mu).
\end{equation}
Under this assumption, the number of parameters can be effectively 
reduced and we can take
$q_I, q_{\romth}$ and ${g_U^2\over M_0}u^2$ as the remaining parameters.
Every oscillation parameter can be written down by using them.  
Taking account of this fact, we find that the atmospheric neutrinos require
$0.017~{\rm eV}~{^<_\sim}~{g_U^2\over M_0}u^2 ~{^<_\sim}~0.023~{\rm eV}$.
This means that $u$ should be $60-70$~keV if we take $M_0\sim 100$~TeV and 
$g_U\sim 0.72$, 
for example. The solar neutrinos impose $(q_I, q_{\romth})$ to be in a certain 
region, which is shown in Fig.1.  From this figure we find that the LMA 
solution is allowed in the wider region than other solutions.
Here it is a crucial problem whether the gauge symmetry including 
U(1)$_X$ can be anomaly free 
in a consistent way 
with the above required region of $(q_I, q_{\romth})$.
On this point, we find that the model can be consistent, although we need to 
introduce additional
chiral superfields such as ${\bf 2}_0, {\bf 1}_{\pm 1}$ and $4({\bf 1}_0)$ of 
SU(2)$_L\times$ U(1)$_Y$, for example.
If we impose the anomaly free conditions and also the massive conditions
for the chiral superfields ${\bf 2}_0, {\bf 1}_{\pm 1}$ at the TeV
region which seems to be required from experiments,  
$q_I$ and $q_{\romth}$ are found 
to satisfy a certain relation as shown in Fig.1.
From this analysis we find that this kind of model can be easily 
constructed in a consistent way.

(2)~$Z^\prime$ nonuniversal coupling and FCNC\\
We assume that U(1)$_X$ has the flavor diagonal but generation dependent
interactions. In general, such kind of interactions can induce dangerous 
FCNC processes, for example, a coherent $\mu$-$e$ conversion,
$\tau\rightarrow 3e,3\mu$, $\mu\rightarrow e\gamma$ and so on.
The additional new $Z^\prime$ interactions can be written in mass
eigenstates as \cite{fcnc}
\begin{eqnarray}
&&{\cal L}_{Z^\prime}=-g_1\left({g_X\over g_1}\cos\xi J^\mu_{(X)}
-\sin\xi J^\mu_{(1)}\right)Z^\prime_\mu, \\
&&J^\mu_{(X)}=\sum_{i,j}\left(\bar\nu_{Li}B_{ij}^{\nu_L}\nu_{Lj}+
\bar\ell_{Li}B_{ij}^{\ell_L}\ell_{Lj}+\bar\ell_{Ri}
B^{\ell_R}_{ij}\ell_R\right),\\
&&B_{ij}^\psi=V^{\psi\dagger} {\rm diag}(q_{\romth}, q_I, q_I)V^\psi,
\end{eqnarray}
where $\ell_L$ and $\ell_R$ stand for the charged leptons. 
$V^\psi$ is a diagonalization matrix of the charged lepton mass 
matrix and it is defined as
\begin{equation}
V^{\psi\dagger}{\cal M}_\psi V^\psi={\rm diagonal}, \qquad
{\cal M}_\psi=\left\{\begin{array}{ll}
m_D^\dagger m_D & (\psi=\ell_R) \\
m_Dm_D^\dagger & (\psi=\ell_L)  \\ \end{array}\right.,
\end{equation}
where $m_D$ is a Dirac mass matrix defined by $\bar\ell_Lm_D\ell_R$.
Because of the U(1)$_X$ constraint, we find that 
$B_{ij}^{\ell_{L,R}}\propto \delta_{ij}$. So nonuniversal coupling of
U(1)$_X$ induces no serious problem in this model.
If the usual conditions $m_{Z^\prime}~{^>_\sim}~600$~GeV 
and $\xi~{^<_\sim}~10^{-3}$ are
satisfied, no contradiction to the present experimental data appears.

(3)~Sneutrino VEVs\\   
Small but nonzero VEVs of sneutrinos $\langle\tilde\nu_\alpha\rangle$
are crucial in this model. If bilinear $R$-parity violating 
terms $\epsilon L_\alpha H_2$ exist in the superpotential, the 
minimization of the scalar potential results in 
\begin{equation}
u\sim\epsilon{\mu\langle H_1\rangle+B_\epsilon\langle H_2\rangle
\over \tilde{m}^2},
\end{equation} 
where $B_\epsilon$ is a soft breaking parameter for $\epsilon L_\alpha
H_2$ and $\tilde m^2$ is an averaged value of the squared 
soft scalar masses of Higgs
scalars. This implies that $u$ can be very small for the small $\epsilon$
satisfying $\epsilon\ll\mu,m,B_\epsilon $. 
The problem is how we can explain the smallness 
of $\epsilon$. Several possible solutions can be considered as in the case of 
$\mu$-term \cite{gm}.
(i)~Giudice-Masiero mechanism~:~ If K\"ahler potential 
includes a term
$ZL_\alpha H_2$ and the supersymmetry is broken by an $F$-term of 
the chiral superfield $Z$, for
example, we have a small one such as $\epsilon\sim {F_Z\over M_{\rm pl}}$. 
(ii)~Higher dimensional term in the superpotential~:~We consider that 
the superpotential includes a nonrenormalizable term such as 
${\bar NN\over M_{\rm pl}^2}SL_\alpha H_2$. If scalar components of 
$\bar N$ and $N$ get the VEVs of the intermediate scale 
through a $D$-flat direction and a scalar component of $S$ gets a TeV 
scale VEV, for example,
we can have $\epsilon\sim{\langle N\rangle^2\over M_{\rm pl}^2}\langle
S\rangle$ which can be sufficiently small.
(iii)~Coupling to the superconformal sector~:~If the MSSM fields couple
to the superconformal sector, they get large anomalous dimensions \cite{ns}.
This can be applied to the explanation of the hierarchy among Yukawa
couplings and the universality of soft scalar masses \cite{ns,yukawa}. 
This idea is also
applicable to the explanation of the smallness of $\epsilon$. For example,
we consider superpotentials
$W_{\rm sc}=y_1\tilde H_1\bar TS+y_2\tilde H_2T\bar S$ and
$W_\mu=\epsilon_\alpha L_\alpha\tilde H_2+\mu_1\tilde
H_1H_2+\mu_2H_1\tilde H_2$, where $T,~\bar{T}$ and $S,~\bar{S}$ are 
chiral superfields in the
superconformal sector and $\tilde H_{1,2}$ are the extended Higgs
chiral superfields. As a result of the couplings with the superconformal 
sector, $\tilde H_{1,2}$ can have a large anomalous 
dimension $\gamma$ and the bilinear $R$-parity violating parameters
$\epsilon_\alpha$ behave as
$\epsilon_\alpha(\Lambda)=\epsilon_\alpha(\Lambda_0)
\left({\Lambda\over\Lambda_0}\right)^\gamma$ \cite{ks}. We find that there is
a large suppression in the case of $\Lambda \ll\Lambda_0$ even if
$\epsilon(\Lambda_0)$ is a weak scale.
Anyway we have many candidates for the mechanism to realize the small 
bilinear $R$-parity violating terms and we can expect that they are
applicable to our model.   
  
\section{An extension of the model}
We consider here an extension of our simple model to include the quark
sector.
For this purpose we introduce two Abelian symmetries
U(1)$_F\times$U(1)$_X$ \cite{origin}.
Although U(1)$_X$ is assumed to be unbroken at a TeV region which is the 
same as in the
previous simple model, we take U(1)$_F$ to be an anomalous U(1) symmetry
which is broken near the Planck scale. It is used for the Froggatt-Nielsen
mechanism to generate the quark and lepton mass hierarchy.
The field contents in this extended models are summarized as follows:
\begin{equation}\begin{array}{lcc}
   &  U(1)_F   &  U(1)_X \\
{\bf 10}_f\equiv(q,u^c,d)_f & (3,2,0) & (\alpha,\alpha,\alpha) \\
{\bf 5}_f^\ast\equiv(d^c, \ell) & (1,0,0) & (q_1,q_1,q_2) \\
{\bf 5}^a\equiv(D,H_2)^a & ((0,0),(0,0)) &((x,z),(p,r)) \\
{\bf 5}^{\ast a}\equiv(\bar D,H_1)^a & ((0,0),(0,0)) &((y,w),(q,s)) \\
{\bf 1}_0 \equiv S_0 & (-1) & (0) \\
{\bf 1}_i \equiv S_i & (0,\cdots,0) & (Q_1,\cdots,Q_6) \\
\end{array}
\end{equation}
where we use the SU(5) representation for the description of the quarks
and the leptons.
We introduce a pair of Higgs doublet chiral superfields only to
guarantee the SM gauge couplings unification additionally to the above list
of field contents. They have no affect to the following results.
Here we impose several phenomenological conditions, that is, 
the gauge anomaly cancellation, 
the proton stability by prohibitting dangerous couplings of the extra 
colored triplets to the SM fields, and also the existence of Higgs
mixings such as  
\begin{equation}
(H_2^1, H_2^2)\left(\begin{array}{cc}
\kappa_1\langle S_1\rangle &\kappa_2\langle S_2\rangle \\ 
\kappa_4\langle S_4\rangle &\kappa_5\langle S_5\rangle \\ 
\end{array}\right)\left(\begin{array}{c} H_1^1 \\ H_1^2\\ \end{array}
\right).
\label{eqd}
\end{equation}
The last condition comes from the requirement to make only one pair of
Higgs doublets massless.
These requirements bring the constraints on the U(1)$_X$ charges 
introduced above. We must study neutrino masses under these constraints. 
If $\kappa_5\langle S_5\rangle/\kappa_4\langle S_4\rangle$ is equal to
$-\kappa_1\langle S_1\rangle/\kappa_2\langle S_2\rangle$ in
eq.~(\ref{eqd}), we find that
$(H_2^1, H_1^\ell\equiv\sin\zeta H_1^1+\cos\zeta H_1^2)$ can play a role 
of the usual Higgs fields, where the mixing angle $\zeta$ is defined as 
$\tan\zeta=\kappa_1\langle S_1\rangle/\kappa_2\langle S_2\rangle$.

Now we discuss the masses and the flavor mixings in the quark 
and lepton sectors. We assume that U(1)$_F$ is broken by the VEV of the
scalar component of $S_0$ which takes a value nearly equal to $M_{\rm pl}$.
In the quark sector U(1)$_F$ controls the flavor mixing by regulating
the number of $S_0$ contained in each nonrenormalizable terms through
the so-called Froggatt-Nielsen mechanism \cite{fn}. 
If we introduce a parameter
$\lambda\equiv\langle S_0\rangle/M_{\rm pl}$, the quark mass matrix can be
written as 
\begin{equation}
M_u\sim\left(\begin{array}{ccc}
\lambda^6 & \lambda^5 & \lambda^3 \\
\lambda^5 & \lambda^4 & \lambda^2 \\
\lambda^3 & \lambda^2 & \ 1 \\ \end{array}\right)\langle H_2^1\rangle, \qquad
M_d\sim\left(\begin{array}{ccc}
\lambda^4\sin\zeta & \lambda^3\sin\zeta & \lambda\sin\zeta \\
\lambda^3\sin\zeta & \lambda^2\sin\zeta & \sin\zeta \\
\lambda^3\cos\zeta & \lambda^2\cos\zeta & \ \cos\zeta \\ 
\end{array}\right)\langle H_1^\ell\rangle.
\end{equation}
From these mass matrices we can obtain the mass eigenvalues and the CKM mixing
as follows,
\begin{eqnarray}
&& m_u : m_c : m_t = \lambda^6 : \lambda^4 : 1, \nonumber \\
&& m_d : m_s : m_b = \lambda^4\sin\zeta : \lambda^2\cos\zeta : \cos\zeta, 
\nonumber \\
&& V_{us} \sim \lambda, \qquad V_{ub}\sim \lambda^3, 
\qquad V_{cb}\sim \lambda^2. 
\end{eqnarray}
This seems to have qualitatively nice features if we take $\lambda\sim
0.22$.
In the lepton sector the charged lepton mass matrix has the same form as
$M_d^T$ because of the SU(5) relation and then we obtain
\begin{equation}
m_e : m_\mu : m_\tau = \lambda^4\sin\zeta : \lambda^2\cos\zeta : \cos\zeta.
\end{equation}
A diagonalization matrix $\tilde U_\ell$ of the charged lepton mass
matrix is found to be 
\begin{equation}
\tilde U_\ell=\left(\begin{array}{ccc}
1 & 0 & \lambda\sin\zeta \\ -\lambda\sin^2\zeta & \cos\zeta & \sin\zeta \\
-\lambda\sin\zeta\cos\zeta & -\sin\zeta & \cos\zeta \\
\end{array}\right).
\end{equation}
\input epsf 
\begin{figure}[tb]
\begin{center}
\epsfxsize=7.6cm
\leavevmode
\epsfbox{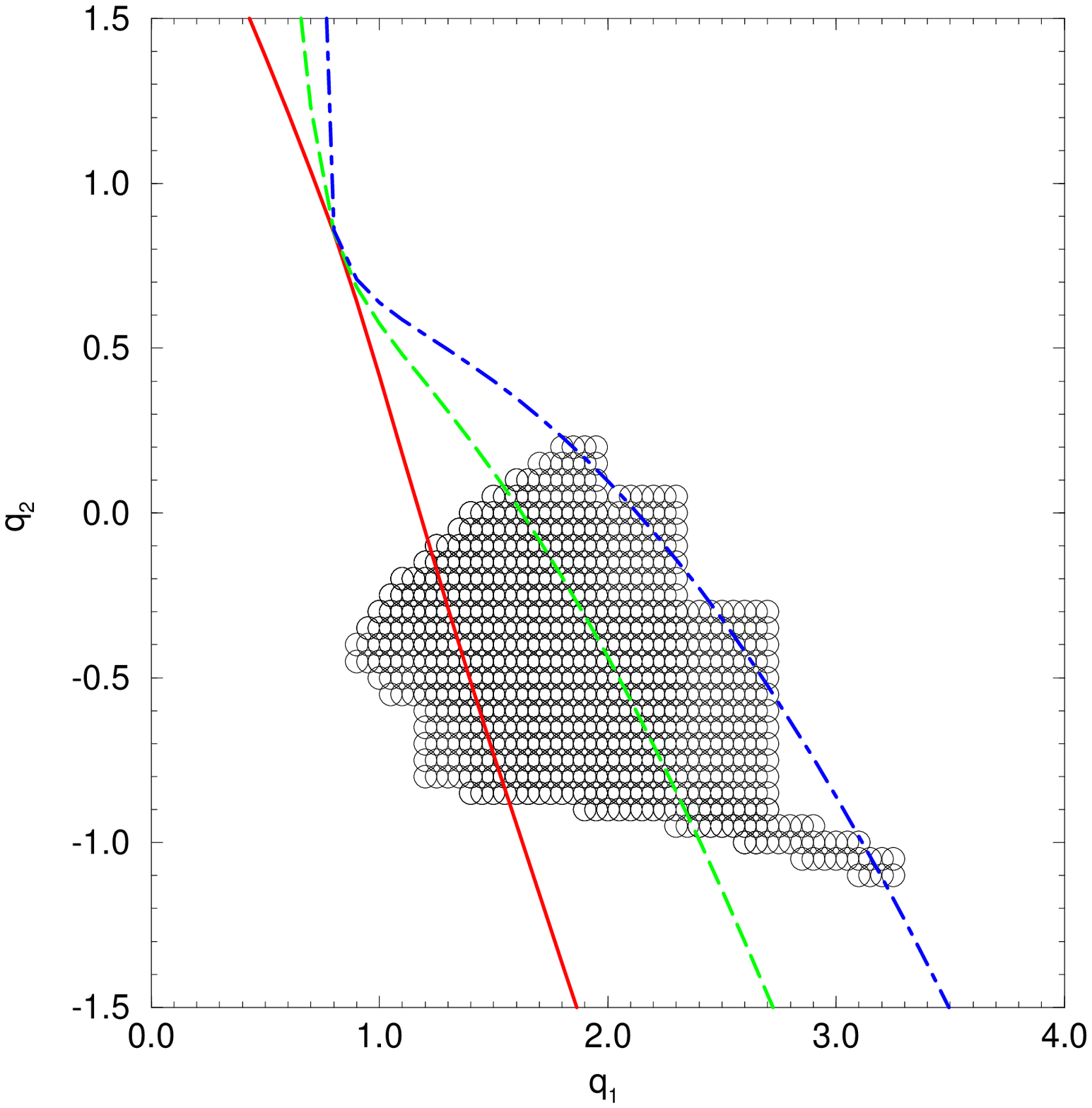}
\end{center}
\vspace*{-1cm}
{\footnotesize Fig. 2~~\  Scatter plots of the U(1)$_X$-charge
of neutrinos which can give the solution to the atmospheric and solar
neutrino problems. Solid, dashed and dot-dashed lines are the relation
between $q_1$ and $q_2$ required by the various phenomenological
conditions. They correspond to $x=1.5,~3$ and 4.5, respectively.}
\end{figure}
For the neutrino sector we can directly use the result in 
the previous simple model. Then, for example, if we take
$\lambda\sim 0.22$, $\cos\theta\sim 1$ and
$\sin\zeta\sim\cos\zeta\sim{1\over\sqrt 2}$, we find that the solar and
atmospheric neutrinos can be explained simultaneously and the 
LMA solution can be applied to the solar neutrino.
The neutrino mass spectrum has a normal hierarchy in this case and
the MNS-matrix can be written as 
\begin{equation}
U^{\rm MNS}=\tilde U_\ell^\dagger U_\nu\sim\left(\begin{array}{ccc}
{1\over\sqrt 2} & -{1\over\sqrt 2} & -{\lambda\over 2} \\ 
{1\over 2} & {1\over 2} & -{1\over\sqrt 2} \\
{1\over 2} & {1\over 2} & {1\over\sqrt 2} \\
\end{array}\right).
\end{equation}
Although the CHOOZ constraint is satisfied \cite{ue3}, 
the maximal mixing is not
realized for the solar neutrino and we have a rather large value of 
$U^{\rm MNS}_{e3}$ such as $-{\lambda\over 2}\simeq -0.11$. 
This is an interesting
prediction of this model.
Neutrinoless double $\beta$-decay seems difficult to be found
in the near future experiments since $\vert M_{ee}\vert\sim {1\over
2}\sqrt{\Delta m^2_{\rm solar}}+{\lambda^2\over 4}\sqrt{\Delta m^2_{\rm
atm}}$ is so small. On the model parameters, if we take $M_0\sim
100$~TeV and $g_U\sim 0.72$, $u$ should be $0.76-1.6$~MeV. The U(1)$_X$
charges $(q_1,q_2)$ should be in the region shown in Fig.2, in which
we also draw the lines which represent a relation between $q_1$ and
$q_2$ required by the previously mentioned various constraints.
Because of the flavor dependent interactions of U(1)$_X$,
FCNC seems to give severe constraints on the model. However, if 
$m_{Z^\prime}~{^>_\sim}~100$~TeV and $\xi~{^<_\sim}~10^{-6}$ are
satisfied, the present experimental bounds for the FCNC can be satisfied. 

\section{Summary}
In this talk we have discussed that the supersymmetry and the neutrino 
masses can be closely related by considering the $R$-parity violation.
If we introduce a generation dependent extra U(1) gauge symmetry at a
TeV region, the masses and mixings of neutrinos can be explained at a 
tree level
due to the mixing among neutrinos and gauginos. Since the supersymmetry is a
promising ingredient to solve the gauge hierarchy problem in the SM, it seems 
to be a very interesting possibility to explain the neutrino masses and
mixings in relation to the supersymmetry.

Our model can be extended so as to include the quark sector
straightforwardly. If we consider two Abelian flavor symmetries
U(1)$_F\times$ U(1)$_X$, the mass and mixing in both quark and lepton
sectors seem to be successfully explained by both the Froggatt-Nielsen
mechanism and the neutrino-gaugino mixing.

\vspace*{5mm}
\noindent
{\Large\bf Acknowledgement}\\

I would like to thank George Zoupanos and organizers of the Summer
Institute CORFU 2001 for their kind hospitality.
This work is partially supported by a Grant-in-Aid for Scientific
Research (C) from Japan Society for Promotion of Science (No.11640267).
\vspace*{5mm}

\end{document}